\documentstyle[12pt]{article}
\begin{document}
\tolerance=5000
\def\be{\begin{equation}}
\def\ee{\end{equation}}
\def\bea{\begin{eqnarray}}
\def\eea{\end{eqnarray}}
\def\nn{\nonumber \\}
\def\cF{{\cal F}}
\def\det{{\rm det\,}}
\def\Tr{{\rm Tr\,}}
\def\e{{\rm e}}
\def\etal{{\it et al.}}
\def\erp2{{\rm e}^{2\rho}}
\def\erm2{{\rm e}^{-2\rho}}
\def\er4{{\rm e}^{4\rho}}
\def\etal{{\it et al.}}

\  \hfill 
\begin{minipage}{3.5cm}
OCHA-PP-106 \\
NDA-FP-40 \\
November 1997 \\
\end{minipage}

\ 

\vfill

\begin{center}

{\large\bf Four-dimensional cosmology  from dilaton 
coupled quantum matter in two dimensions}

\vfill

{\sc Tomoko KADOYOSHI}\footnote{e-mail: 
kado@fs.cc.ocha.ac.jp} \\
{\sc Shin'ichi NOJIRI$^{\clubsuit}$}\footnote{
e-mail : nojiri@cc.nda.ac.jp} and 
{\sc Sergei D. ODINTSOV$^{\spadesuit}$}\footnote{
e-mail : odintsov@quantum.univalle.edu.co, 
odintsov@kakuri2-pc.phys.sci.hiroshima-u.ac.jp}

\vfill

{\sl Department of Physics, 
Ochanomizu University \\
Otsuka, Bunkyou-ku Tokyo 112, JAPAN}

\ 

{\sl $\clubsuit$ 
Department of Mathematics and Physics \\
National Defence Academy, 
Hashirimizu Yokosuka 239, JAPAN}

\ 

{\sl $\spadesuit$ 
Tomsk Pedagogical University, 634041 Tomsk, RUSSIA \\
and \\
Dep.de Fisica, Universidad del Valle, 
AA25360, Cali, COLOMBIA \\
}

\vfill

{\bf ABSTRACT}

\end{center}

The reduction of 4D Einstein gravity with $N$ minimal 
scalars leads to specific 2D dilaton gravity with dilaton coupled scalars. 
Applying $s$-wave and large $N$ approximation (where large $N$ quantum 
contribution due to dilaton itself is taken into account) we 
study 2D cosmology for CGHS model and for reduced Einstein gravity 
(in both cases with dilaton coupled scalars). Numerical study shows 
that in most cases we get 2D singular Universe which may correspond to
big bang. Nevertheless, big crunch or non-singular Universes 
are also possible. For reduced 4D Einstein gravity one can regard 
the obtained 2D cosmology with time-dependent dilaton as
4D Kantowski-Sacks Universe of singular,non-singular or big crunch type.

\ 

\noindent
PACS: 

\newpage

\noindent
{\bf 1. Introduction.}
It is quite well-known fact that models of two-dimensional 
dilaton gravity with matter may describe qualitatively the main 
properties of Hawking gravitational collapse \cite{I} taking 
account of quantum matter back effects \cite{II,III} (for a review, 
see \cite{IV} and for a very incomplete list of earlier related 
works, see \cite{V}). These models attract the interest being 
often exactly solvable ones and representing toy laboratory for study of Hawking radiation. From another point, the action of 
ref.\cite{II}:
\be
\label{1}
S=\int d^2 x \sqrt{-g} \left[-{1 \over 16\pi G}\e^{-2\phi}
\left( R+4(\nabla \phi)^2 + 4\lambda^2\right) 
 + {1 \over 2}\sum_{i=1}^N(\nabla \chi_i)^2\right]\ .
\ee
which belongs to more general class of renormalizable dilatonic 
gravities \cite{VI} may be relevant to description of radial 
modes of 4D extremal black holes \cite{VII}.

However one can restrict 4D action of Einstein gravity with cosmological term and with matter described by $N$ minimal scalars 
$\chi_i$ to the metrics with spherical symmetry:
\be
\label{2}
ds^2=g_{\mu\nu}dx^\mu dx^\nu + \e^{-2\phi}d\Omega
\ee
Here two-dimensional metric and dilaton depend only from
time and radius.
Then spherically reduced action reads
\be
\label{3}
S_{red}=\int d^2x \sqrt{-g}\e^{-2\phi}
\left[-{1 \over 16\pi G}
\{R + 2(\nabla  \phi)^2 -2\Lambda + 2\e^{2\phi}\} 
+ {1 \over 2}\sum_{i=1}^N(\nabla \chi_i)^2 \right]\ .
\ee
Theory (\ref{3}) represents the model which is completely 
different from (\ref{1}) not only in gravitational sector but 
also in matter sector.
Unlike to the previous models \cite{I}--\cite{V}, one has to 
take into account quantum effects of dilaton coupled scalar 
and not of minimal scalars as in (\ref{1}). 
Note that one could start from more general reduction having an 
arbitrary function $f(\phi)$ instead of $\e^{-2\phi}$ in (\ref{2}).
In this case, one would have more general theory in two dimensions 
with the change $\e^{-2\phi}\rightarrow f(\phi)$ in (\ref{3}) and 
the correspondent change of dilaton dependent kinetic-like term 
in (\ref{3}). Quantum field theory of dilaton coupled 
two-dimensional scalars (and spinors) with arbitrary 
dilaton function in curved spaces with dilaton has been 
first studied in ref.\cite{VIII} where one-loop effective 
action has been found.

It is clear that it is action (\ref{3}) which should be used 
for the study of reduced 4D theory from 2D dimensional point 
of view in $s$-wave and large $N$ approximation but not action 
(\ref{1}). For that purpose, one needs the conformal anomaly 
for dilaton coupled scalars. Such anomaly for dilatonic 
coupling $\e^{-2\phi}$ as well as correspondent anomaly 
induced effective action has been calculated by Bousso and 
Hawking \cite{IX}. For an arbitrary dilatonic coupling $f(\phi)$
the conformal anomaly and anomaly 
induced effective action has been found in ref.\cite{X} 
(actually, this result is presented in non-explicit form 
already in ref.\cite{VIII}).

Hawking radiation for 2D black holes with account of quantum 
effects of dilaton coupled scalars has been investigated in 
refs.\cite{XI,XII}. It is interesting to note in this respect that 
dilaton coupled spinor leads to the same trace anomaly as minimal 
spinor \cite{VIII}. Using this fact, Hawking radiation in 2D dilatonic 
supergravity (for a review, see\cite{XIII})
 has been also discussed \cite{XIV}.

In the recent paper \cite{XV}, quantum evolution of 4D black holes 
has been studied in $s$-wave and 
large N approximation taking account of dilaton coupled quantum matter. The
possibility of anti-evaporation of black holes has been found.

In the present paper, we discuss another class of phenomena where 
quantum effects of dilaton coupled $2D$ matter may be relevant, i.e., 
we discuss quantum cosmological models. 
We concentrate on the spherically symmetric metrics of 
Kantowski-Sachs form \cite{XVI}:
\be
\label{4}
ds^2=g_{\mu\nu}dx^\mu dx^\nu + \e^{-2\phi}d\Omega
\ee
where $g_{\mu\nu}=a^2(t)\eta_{\mu\nu}$. Such metric (\ref{4}) 
describes the Universe with a $S^1\times S^2$ spatial geometry 
(for some its properties, see \cite{XVII}). The back reaction 
problem for the time-dependent metric (\ref{4}) in CGHS model 
with minimally coupled scalars has been studied in refs.\cite{XVIII}.
Here, also 4D interpretation of results obtained has been presented.

However as it follows from above discussion, the correct approach to 
4D quantum cosmology (\ref{4}) in $s$-wave and large $N$ 
approximation is to start from the action (\ref{3}) and take 
into account back-reaction via trace anomaly of dilaton coupled 
scalars. That is the purpose of present paper. 

In order to see the difference from CGHS model with minimal scalars, 
we start in next section from CGHS model with dilaton coupled 
scalars. In this case, the model is not exactly solvable anymore. 
The numerical study of 2D time-dependent cosmology with time 
dependent dilaton is presented in large $N$ approximation. 
The comparison is done with results of refs.\cite{XVIII}. 
Note that such 2D cosmological quantum models may be of interest by 
itself.

In the section 3, we repeat the same study for 4D reduced theory 
with the action (\ref{3}). We again discuss 2D cosmology. However, 
now 4D interpretation as Kantowski-Sacks models may be given.  

\

\noindent{\bf 2. CGHS model with dilaton coupled matter}
The starting point is Eq.(\ref{1}). 
The large $N$ effective action produced by quantum dilaton coupled 
scalars 
(with account of quantum dilaton contribution)
is \cite{X}
\bea
\label{qc}
W&=&-{1 \over 2}\int d^2x \sqrt{-g} \left[ 
{N \over 48\pi}R{1 \over \Delta}R 
-{1 \over 8\pi}\left({{f'}^2 \over f} - f''\right)
(\nabla^\lambda \chi_i)(\nabla_\lambda \chi_i) 
{1 \over \Delta}R \right. \nn
&& \left. -{N \over 16\pi}{{f'}^2 \over f^2}
\nabla^\lambda \phi
\nabla_\lambda \phi {1 \over \Delta}R 
-{N \over 8\pi}\ln f R \right]\ .
\eea
In the conformal gauge
\be
\label{cg}
g_{\pm\mp}=-{1 \over 2}\e^{2\rho}\ ,\ \ 
g_{\pm\pm}=0
\ee
the equations of motion which follow from Eq.(\ref{1}) 
 (where last ter is multiplied by $f(\phi)$) plus Eq.(\ref{qc}) 
are obtained by the variation 
over $g^{\pm\pm}$, $g^{\pm\mp}$, $\phi$, $\chi$  
\bea
\label{eqnpp}
0&=&\e^{-2\phi}\left(4\partial_\pm \rho
\partial_\pm\phi - 2 \left(\partial_\pm\phi\right)^2 
\right) + {f \over 2}\sum_i(\partial_\pm\chi_i)^2
+{N \over 12}\left( \partial_\pm^2 \rho 
- \partial_\pm\rho \partial_\pm\rho \right) \nn
&& +{N \over 8} \left\{\left( 
\partial_\pm \tilde\phi \partial_\pm\tilde\phi \right)
\rho+{1 \over 2}{\partial_\pm \over \partial_\mp}
\left( \partial_\pm\tilde\phi 
\partial_\mp\tilde\phi \right)\right\} 
+{N \over 8}\left\{ 
-2 \partial_\pm \rho \partial_\pm \tilde\phi 
+\partial_\pm^2 \tilde\phi \right\} \nn
&& +{1 \over 4}\sum_i\left\{g(\partial_\pm\chi_i)^2
\rho+{1 \over 2}{\partial_\pm \over \partial_\mp}
\left( g\partial_\pm\chi_i \partial_\mp\chi_i 
\right)\right\} + t^\pm(x^\pm) \\
\label{req}
0&=&\e^{-2\phi}\left(2\partial_+
\partial_- \phi -4 \partial_+\phi\partial_-\phi 
- \lambda^2 \e^{2\rho}\right) \nn
&& -{N \over 12}\partial_+\partial_- \rho
-{N \over 16}\partial_+ \tilde\phi 
\partial_- \tilde \phi 
-{N \over 8}\partial_+\partial_-\tilde\phi 
-{1 \over 8}g\sum_i\partial_+\chi_i\partial_-\chi_i \\
\label{eqtp}
0&=& \e^{-2\phi}\left(-4\partial_+
\partial_- \phi +4 \partial_+\phi\partial_-\phi 
+2\partial_+ \partial_- \rho
+ \lambda^2 \e^{2\rho}\right) 
+ {1 \over 4}f'\sum_i\partial_+\chi_i\partial_-\chi_i \nn
&& -N\tilde\phi'\left\{
{1 \over 16}\partial_+(\rho \partial_-\tilde\phi)
+{1 \over 16}\partial_-(\rho \partial_+\tilde\phi)
-{1 \over 8}\partial_+\partial_-\rho \right\} \nn
&& +{1 \over 8}g'\rho\sum_i\partial_+\chi_i\partial_-\chi_i \\
\label{eqchi}
0&=&\partial_+(f\partial_-\chi_i)
+\partial_-(f\partial_+\chi_i)
-{1 \over 2}\left\{\partial_+(g\rho\partial_-\chi_i)
+\partial_-(g\rho\partial_+\chi_i)
\right\}\ .
\eea
Here $\tilde\phi = \ln f$, $g={{f'}^2 \over f}-f''$ 
and $t^\pm(x^\pm)$ is a function which is determined by 
the boundary condition.
When we consider the cosmological problem,
we can assume that all the fields depend only on 
time $t$ and replace 
$\partial_\pm\rightarrow {1 \over 2}\partial_t$.
Then Eq.(\ref{eqnpp}) tells $t^\pm$ is a constant:
$t^\pm={N \over 4}t_0$ and we obtain
\bea
\label{eqnpp2}
0&=&\e^{-2\phi}\left(4\partial_t \rho
\partial_t\phi - 2 \left(\partial_t\phi\right)^2 
\right) + {f \over 2}\sum_i(\partial_t\chi_i)^2\nn
&& +{N \over 12}\left( \partial_t^2 \rho 
- \partial_t\rho \partial_t\rho \right) 
+{N \over 8} \left(\rho+{1 \over 2}\right) 
\partial_t\tilde\phi \partial_t\tilde\phi \nn
&& +{N \over 8}\left\{ 
-2 \partial_t \rho \partial_t \tilde\phi 
+\partial_t^2 \tilde\phi \right\} 
+{1 \over 4}\left(\rho + {1 \over 2}\right)\sum_i
g(\partial_t\chi_i)^2 + N t_0 \\
\label{req2}
0&=&\e^{-2\phi}\left(2\partial_t
\partial_t \phi -4 \partial_t\phi\partial_t\phi 
- 4\lambda^2 \e^{2\rho}\right) \nn
&& -{N \over 12}\partial_t^2 \rho
-{N \over 16}\partial_t \tilde\phi 
\partial_t \tilde \phi 
-{N \over 8}\partial_t^2\tilde\phi 
-{1 \over 8}g\sum_i\partial_t\chi_i\partial_t\chi_i \\
\label{eqtp2}
0&=& \e^{-2\phi}\left(-4\partial_t
\partial_t \phi +4 \partial_t\phi\partial_t\phi 
+2\partial_t \partial_t \rho
+ 4\lambda^2 \e^{2\rho}\right) 
+ {1 \over 4}f'\sum_i\partial_t\chi_i\partial_t\chi_i \nn
&& - {N \over 8}\varphi' \left\{
\partial_t(\rho \partial_t\tilde\phi)
-\partial_t^2\rho \right\} 
+{1 \over 8}g'\rho\sum_i\partial_t\chi_i\partial_t\chi_i \\
\label{eqchi2}
0&=&\partial_t\left\{\left(f-{1 \over 2}g\rho\right)
\partial_t\chi_i\right\} \ .
\eea
Eq.(\ref{eqchi2}) can be integrated to be
$\left(f-{1 \over 2}g\rho\right)\partial_t\chi_i=a_i$.
Here $a_i$ is a constant of the integration.

When we investigate the cosmological problem, 
it is often more convenient to use the 
gauge where $g_{tt}=-1$ and $g_{tx}=g_{xt}=0$ 
than to use the conformal gauge (\ref{cg}).
Since the metric under consideration 
depend only on $t$, the change of the gauge choice from 
conformal gauge is given by $d{\hat t} = \e^\rho dt$, 
accordingly
\be
\label{trp}
\partial_t = \e^\rho\partial_{\hat t}\ ,\ \ \ 
\partial_t^2 = \e^{2\rho}\left(
\partial_{\hat t}^2 + (\partial_{\hat t}\rho)
\partial_{\hat t}\right)\ .
\ee
In the following we call $t$ as the conformal 
time and ${\hat t}$ as the cosmological time.
   
For $f(\phi)=\e^{-2\phi}$  $(\tilde\phi=-2\phi)$,
we find $g=0$ 
and the equations are reduced to be
\bea
\label{eqnpp3}
0&=&\e^{-2\phi}\left(4\partial_t \rho
\partial_t\phi - 2 \left(\partial_t\phi\right)^2 
+ {1 \over 2}\sum_i(\partial_t\chi_i)^2\right)
+{N \over 12}\left( \partial_t^2 \rho 
- (\partial_t\rho)^2 \right) \nn
&&+{N \over 2} \left(\rho+{1 \over 2}\right) 
(\partial_t\phi)^2 
+{N \over 4}\left\{ 2 \partial_t \rho \partial_t \phi 
-\partial_t^2 \phi \right\} + N t_0 \\
\label{req3}
0&=&\e^{-2\phi}\left(2\partial_t^2 \phi -4 (\partial_t\phi)^2 
- 4\lambda^2 \e^{2\rho}\right) 
-{N \over 12}\partial_t^2 \rho
-{N \over 4}(\partial_t \phi)^2 
+{N \over 4}\partial_t^2 \phi \\
\label{eqtp3}
0&=& \e^{-2\phi}\left(-4\partial_t^2 \phi 
+4 (\partial_t\phi)^2 
+2\partial_t^2 \rho
+ 4\lambda^2 \e^{2\rho}
- {1 \over 2}\sum_i(\partial_t\chi_i)^2
\right) \nn
&& + {N \over 4} \left\{
-2\partial_t(\rho \partial_t\phi)
-\partial_t^2\rho \right\}  \\
\label{eqchi3}
0&=&\partial_t\left\{
\e^{-2\phi}\partial_t\chi_i\right\} \ .
\eea
and
\be
\label{pchi2}
\partial_t\chi_i=a_i\e^{2\phi}
\ee

First we consider $N\rightarrow\infty$ case where 
the equations become simpler:
\bea
\label{eqnpp4}
0&=&{a^2 \over 2}\e^{2\phi} 
+{1 \over 12}\left( \partial_t^2 \rho 
- (\partial_t\rho)^2 \right) 
+{1 \over 2} \left(\rho+{1 \over 2}\right) 
(\partial_t\phi)^2 \nn
&& +{1 \over 4}\left\{ 2 \partial_t \rho \partial_t \phi 
-\partial_t^2 \phi \right\} + t_0 \\
\label{req4}
0&=& -{1 \over 12}\partial_t^2 \rho
-{1 \over 4}(\partial_t \phi)^2 
+{1 \over 4}\partial_t^2 \phi \\
\label{eqtp4}
0&=& {a^2 \over 2}\e^{2\phi} 
+ {1 \over 4} \left\{
-2\partial_t(\rho \partial_t\phi)
-\partial_t^2\rho \right\}  \ .
\eea
Here we used Eq.(\ref{pchi2}) and 
$a^2\equiv {1 \over N}\sum_i a_i^2$.
Combining (\ref{eqnpp4}) and (\ref{req4}), we obtain
\be
\label{cmeq}
0=-{1 \over 12}(\partial_t\rho)^2 
+ {1 \over 2}\rho(\partial_t\phi)^2 
+{1 \over 2}\partial_t\rho\partial_t\phi + t_0
+{a^2 \over 2}\e^{2\phi}\ .
\ee
When $t_0=a=0$, we find
\be
\label{rhophi}
\partial_t\phi={1 \over 2\rho}
\left\{-1 \pm \sqrt{1+{2 \over 3}\rho}\right\}\partial_t\rho
\ee
that is,
\be
\label{rhophi2}
\phi=\int {d\rho \over 2\rho}
\left\{-1 \pm \sqrt{1+{2 \over 3}\rho}\right\}\ .
\ee
Substituting Eq.(\ref{rhophi}) into Eq.(\ref{eqtp4}), 
we obtain
\be
\label{Nrho}
\rho=-{3 \over 2} + \left(\alpha t + \beta
\right)^{2 \over 3}\ .
\ee
Here $\alpha$ and $\beta$ are the constant of the 
integration.
The solution (\ref{Nrho}) has the curvature singularity 
when $\alpha t + \beta=0$. 
Since $\rho$ is monotonically increasing function 
with respect to $t$ (and also the 
cosmological time ${\hat t}$ since $\e^\rho$ is 
positive) when $\alpha>0$, the solution (\ref{Nrho}) 
can express the expanding universe solution where 
the big bang is given by the curvature singularity 
at $\alpha t + \beta=0$.

We now consider Eqs.(\ref{eqnpp4}), (\ref{req4}) 
and (\ref{eqtp4}) numerically.
If we define new variables $P$ and $R$ by 
\be
\label{PR}
P=\partial_t \phi\ ,\ \ R=\partial_t \rho\ ,
\ee
Eqs. (\ref{eqnpp4}) and (\ref{req4}) give
\bea
\label{Eq1}
\partial_t P &=& \left(1 + {2 \over 3}\rho\right)^{-1}
\left(P^2 - {2 \over 3}RP 
- {2 \over 3} a^2 \e^{2\phi} \right) \\
\label{Eq2}
\partial_t R &=& -2\left(1 + {2 \over 3}\rho\right)^{-1}
\left(\rho P^2 + RP + a^2\e^{2\phi}\right)
\eea
and Eq.(\ref{cmeq}) can be rewritten as 
\be
\label{inicon}
0=Q\equiv -{1 \over 6}R^2 
+ \rho P^2 + RP +2t_0 + a^2 \e^{2\phi}\ .
\ee
Using Eq.(\ref{inicon}) we can rewrite Eq.(\ref{Eq2}) 
as follows:
\be
\label{Eq2b}
\partial_t R = -\left(1 + {2 \over 3}\rho\right)^{-1}
\left({1 \over 3}R^2 - 4t_0 \right)\ .
\ee
Eqs.(\ref{Eq1}) and (\ref{Eq2}) tell that there is a 
singularity when
$\rho=\rho_c\equiv -{3 \over 2}$.
A special case of the singularity is that of (\ref{Nrho}).
Therefore this singularity can be interpreted to be 
the singularity of the big-bang.

Eq.(\ref{inicon}) can be regarded as the constraint 
for the initial values of $\phi$, $\rho$, $P$ and $R$. 
We can directly check that $\partial_t Q=0$ by using 
(\ref{Eq1}) and (\ref{Eq2}).
Then we solve the equations (\ref{PR}),  
(\ref{Eq1}) and (\ref{Eq2}) choosing the initial 
values of $\phi$, $\rho$, $P$ and $R$ satisfying 
Eq.(\ref{inicon}). 
Note that also Eq.(\ref{Eq2b}) does not contain $\phi$ 
but only $\rho$. 

Therefore if we are interested in the behavior of $\rho$, 
that is, the time-development of the universe, we only need 
to solve Eq.(\ref{Eq2b}).
If we define a new variable $\hat\rho$ by
$\hat\rho=\left(\rho + {3 \over 2}\right)^{3 \over 2}$,
we can rewrite (\ref{Eq2b}) as follows
\be
\label{hatrhoeq}
\partial_t^2\hat\rho=9t_0\hat\rho^{-{1 \over 3}}\ .
\ee
Eq.(\ref{hatrhoeq}) can be compared with the Newton's 
equation of motion of a particle with unit mass in the 
potential $V(\hat\rho)$ given by
\be
\label{pot}
V(\hat\rho)=-{27 \over 4}t_0 \hat \rho^{4 \over 3}\ .
\ee
Therefore the expanding universe turns to shrink if $t_0<0$ 
as in the Friedmann universe with $\Omega>0$. 
On the other hand, if 
$t_0>0$, the universe continues to expand as 
in $\Omega<0$.

Numerical study of Eq.(\ref{Eq2b}) is given in 
Figures. 
The vertical line corresponds to $\rho$ 
and the horizontal line to $t$.
Fig.1 represents the case $\rho(0)=-1.4$, $\dot\rho(0)=1$, 
and $t_0=1$ (solid line), $t_0=0$ (dashed line) 
and $t_0=-1$ (dot-dashed line).
In Fig.2, the horizontal line is given by the cosmolgical 
time $\hat t$ in (\ref{trp}). 
Fig.2 also 
represents the case $\rho(0)=-1.4$, $\dot\rho(0)=1$, 
and $t_0=1$ (solid line), $t_0=0$ (dashed line) 
and $t_0=-1$ (dot-dashed line).

We now consider the finite $N$ case.
Using (\ref{eqnpp3}) and (\ref{req3}), we obtain
\bea
\label{eqphi}
\partial_t^2 \phi &=&
-2\left(1+{N \over 8}\e^{2\phi}\right)
\partial_t\rho \partial_t\phi 
+ 3\left(1 - {N \over 12}\e^{2\phi}\rho
\right)(\partial_t\phi)^2 \nn
&& + {N \over 24}\e^{2\phi}(\partial_t\rho)^2 
-{N \over 4}a^2\e^{4\phi} 
+ 2\lambda^2\e^{2\rho} - {1 \over 2}\e^{2\phi}N t_0 \\
\label{eqrho}
\partial_t^2 \rho 
&=& 6\left\{ \left(1 + {4 \over N}\e^{-2\phi}\right) 
-\left(1 + {N \over 8}\e^{2\phi}\right)\rho \right\}
(\partial_t\phi)^2 \nn
&& -{48 \over N}\e^{-2\phi}
\left(1 + {N \over 8}\e^{2\phi}\right)^2 
\partial_t\rho \partial _t \phi 
+ \left(1 
+ {N \over 8}\e^{2\phi}\right)(\partial_t\rho)^2 \nn
&& + 6\lambda^2\e^{2\rho} 
-6 \left(1 + {N \over 8}\e^{2\phi}\right)a^2 \e^{2\phi} 
-12\left(1 + {N \over 8}\e^{2\phi}\right) t_0 \ .
\eea
If we delete the terms in (\ref{eqnpp3}), (\ref{req3}) 
and (\ref{eqtp3}) which contain 
the second order derivatives, we obtain 
\bea
\label{inicon2}
0&=& \left\{ -{96 \over N}\e^{-4\phi}-4 \e^{-2\phi} 
+ 2N + {3 N^2 \over 16}\e^{2\phi} 
+ N\left(1 + {N \over 8}\e^{2\phi}\right)\rho\right\}
\partial_t \rho \partial_t \phi \nn
&& + \left\{ -{3N \over 2} - 2 \e^{-2\phi} 
+ {48 \over N}\e^{-4\phi}  -{1 \over 2}\left(
N + 24\e^{-2\phi} - {3 N^2 \over 8}\e^{2\phi}\right)\rho 
\right. \nn
&& \left. 
+{N^2 \over 8}\e^{2\phi}\rho^2\right\}(\partial_t\phi)^2 
+ \left( -{N \over 6}+2\e^{-2\phi}-{N^2 \over 32}\e^{2\phi} 
- {N^2 \over 48}\e^{2\phi}\rho \right)(\partial_t\rho)^2 \nn 
&& +\left(8\e^{-2 \phi} - N\rho - {3N \over 2}\right)
\lambda^2\e^{2\rho} 
+ 2N\left(1 + {3N \over 16}\e^{2\phi} 
- {12 \over N}\e^{-2\phi}
+ {N \over 8}\rho\e^{2\phi}\right)t_0 \nn
&& + \left({3N \over 4}+2\e^{-2 \phi} 
- {48 \over N}\e^{-4\phi} + {N \over 2}\rho\right)
{Na^2 \over 4}\e^{4\phi}\ .
\eea
Choosing the initial condition which satisfies 
(\ref{inicon2}), we can solve (\ref{eqphi}) and (\ref{eqrho}).
Otherwise (\ref{inicon2}) can be regarded to be the equation 
which determines $t_0$ from the initial value.
Note that when $a=\lambda^2=0$, 
(\ref{eqphi}) and (\ref{eqrho}) have trivial 
solutions where $\rho$ and $\phi$ are constants
($t_0=0$).

We have calculated numericaly several cases 
with the initial condition at $t=0$
\be
\label{inicon3}
\phi=\dot\phi=\rho=\dot\rho=0\ .
\ee
When $\lambda^2<0$, $\phi$ and $\rho$ decrease 
monotonically in most of cases. A typical example 
is given in Fig.3.  $\phi$ is given in solid line and 
$\rho$ in dashed line.
The parameters in Fig.3 are chosen to be 
$N=1$, $a=1$, $\lambda^2=-1$ and $t_0=-0.823699$.
When $\lambda^2>0$, $\phi$ 
increases monotonically and $\rho$ increases first 
and decreases in most of cases. A typical example 
is also given in Fig.4 ($\phi$ is solid line and 
$\rho$ dashed line)
where the parameter are chosen to be 
$N=1$, $a=1$, $\lambda^2=1$ and $t_0=-0.141791$.
In the above two cases of Fig.3 and Fig.4, 
there appear singularities in the 
finite conformal time $t$. 
The singularities occur when $\rho$ is decreasing, what 
tells that the singularities occur in the finite cosmological 
time $\hat t$ in Eq.(\ref{trp}).
Since the system is, 
of course, invariant if we change $t$ with $-t$, 
the singularity can be regarded to express the big bang if 
we reverse the direction of the time $t$. 
Due to the initial condition (\ref{inicon3}), we can paste 
the solutions in Figs.3,4 with the time reversed solution 
at $t=0$. The combined solution would express the process 
that the universe generated by the big bang disappears by the 
big crunch. 
The obtained solution would correspond to $n>1$ case in 
\cite{XVIII}.

We also find a solution without singularity, which is 
given in Fig.5 ($\phi$ is solid line and 
$\rho$ dashed line)
where the parameter are chosen to be 
$N=1$, $a=1$, $\lambda^2=0$ and $t_0=-0.523121$.
In Fig.5, $\phi$ oscillates like sine or cosine function
and $\rho$ decreases slowly with vibration.
Such a vibrating solutions also appear when $N$ is large. 
In Fig.6 ($\phi$ is solid line and 
$\rho$ dashed line)
where the parameters are chosen to be 
$N=100$, $a=1$, $\lambda^2=1$ and $t_0=-0.487264$.
The solution in Fig.6 has a singularity at the finite 
conformal and/or cosmological time.
Any solution which shows the behavior 
like in Fig.5 and Fig.6 does not seen in \cite{XVIII}.

In the above solutions, if we exchange the role of time 
and radius, we obtain a static object, which can be regarded 
as a kind of black hole.

\

\noindent{\bf 3. The reduced model.}
 Two-dimensional dilaton gravity with scalars can be regarded as the 
dimensional reduction of four dimensional Einstein gravity 
whose metric is given by
\be
\label{4dmetric}
ds^2=\e^{2\rho}(-dt^2 + dr^2) 
+ \e^{-2\phi}d\Omega^2 .
\ee
The scalar curvature corresponding to the metric 
(\ref{4dmetric}) is given by
\be
\label{4dR}
R_{4d}=-\e^{-2\rho}\left\{
2\partial^2_t\rho - 4\partial^2_t\phi 
+ 6(\partial_t \phi)^2\right\}\ .
\ee
The obtained reduced Einstein-scalar action has the 
form of Eq.(\ref{3}).
 Adding to it the quantum correction obtained from 
conformal anomaly we may get the following equations 
of motion similar to (\ref{eqphi}) and (\ref{eqrho}):
\bea
\label{reqphi}
\partial_t^2 \phi &=&
-2\left(1+{N \over 8}\e^{2\phi}\right)
\partial_t\rho \partial_t\phi 
+ \left({5 \over 2} - {N \over 4}\e^{2\phi}\rho
\right)(\partial_t\phi)^2 \nn
&& + {N \over 24}\e^{2\phi}(\partial_t\rho)^2 
-{N \over 4}a^2\e^{4\phi} 
+\lambda^2\e^{2\rho} + \e^{2\phi+2\rho} 
- {1 \over 2}\e^{2\phi}N t_0 \\
\label{reqrho}
\partial_t^2 \rho 
&=& 6\left\{ \left({1 \over 4} + {3 \over N}\e^{-2\phi}\right) 
-\left(1 + {N \over 8}\e^{2\phi}\right)\rho \right\}
(\partial_t\phi)^2 \nn
&& -{48 \over N}\e^{-2\phi}
\left(1 + {N \over 8}\e^{2\phi}\right)^2 
\partial_t\rho \partial _t \phi 
+ \left(1 
+ {N \over 8}\e^{2\phi}\right)(\partial_t\rho)^2 \\
&& + 3\lambda^2\e^{2\rho} +3\e^{2\phi+2\rho}
-6 \left(1 + {N \over 8}\e^{2\phi}\right)a^2 \e^{2\phi} 
-12\left(1 + {N \over 8}\e^{2\phi}\right) t_0 \ .\nonumber
\eea
We also obtain the equation for the initial condition 
corresponding to (\ref{inicon2}):
\bea
\label{rinicon2}
0&=& \left\{ -{96 \over N}\e^{-4\phi}-8 \e^{-2\phi} 
+ 2N + {3 N^2 \over 16}\e^{2\phi} 
+ N\left(1 + {N \over 8}\e^{2\phi}\right)\rho\right\}
\partial_t \rho \partial_t \phi \nn
&& + \left\{ -{9N \over 8} +3 \e^{-2\phi} 
+ {24 \over N}\e^{-4\phi}  -{1 \over 2}\left( {3 \over 2}
N + 24\e^{-2\phi} - {3 N^2 \over 8}\e^{2\phi}\right)\rho 
\right. \nn
&& \left. 
+{N^2 \over 8}\e^{2\phi}\rho^2\right\}(\partial_t\phi)^2 
+ \left( -{N \over 12}+2\e^{-2\phi}-{N^2 \over 32}
\e^{2\phi} 
- {N^2 \over 48}\e^{2\phi}\rho \right)(\partial_t\rho)^2 \nn 
&& -\left(-6\e^{-2 \phi} + {N \over 2}\rho 
+ {3N \over 4}\right)\lambda^2\e^{2\rho} 
- \left(-4\e^{-2 \phi} + {N \over 2}\rho 
+ {3N \over 4}\right)\e^{2\phi+2\rho} \nn
&&+ 2N\left({1 \over 2} + {3N \over 8}\e^{2\phi} 
- {12 \over N}\e^{-2\phi}
+ {N \over 16}\rho\e^{2\phi}\right)t_0 \nn
&& + \left({3N \over 4}
- {48 \over N}\e^{-4\phi} + {N \over 2}\rho\right)
{Na^2 \over 4}\e^{4\phi}\ .
\eea
Here $\lambda^2=-\Lambda$.
The behaviors of the numerical solutions of the equations
(\ref{reqphi}), (\ref{reqrho}) and (\ref{rinicon2}) 
with the initial condition (\ref{inicon3}) 
are very similar 
to those of CGHS type model.
When $\lambda^2<0$, $\phi$ and $\rho$ decrease 
monotonically in most of cases as in Fig.3. 
A typical example 
is given in Fig.7 
where the parameters are chosen to be
$N=1$, $a=0$ and $\lambda^2=-1$.
in Fig.7, solid line and dashed line represent $\phi$ 
and $\rho$ respectively and dot-dashed line shows the 
behaviour of  the 4d scalar curvature $R_{4d}$ 
in (\ref{4dR}).
When $\lambda^2>0$, $\phi$ 
increases monotonically and $\rho$ increases first 
and decreases in most of cases. A typical example 
is also given in Fig.8 (solid line, dashed line and dot-dashed 
line represent $\phi$, $\rho$ and $R_{4d}$, respectively)
where the parameters are chosen to be 
$N=1$, $a=0$, $\lambda^2=1$.
In the above two cases of Fig.7 and Fig.8, 
there also appear singularities in the 
finite conformal time $t$ and/or in the finite cosmological 
time $\hat t$.
Vibrating solutions also appear when $N$ is large. 
In Fig.9 (solid line, dashed line and dot-dashed 
line represent $\phi$, $\rho$ and $R_{4d}$, respectively) 
where the parameters are chosen to be 
$N=100$, $a=1$ and $\lambda^2=1$.
The solution in Fig.9 has a singularity at the finite 
conformal and/or cosmological time.
Since $\rho$ becomes small at the singularity, 
the singularity 
would correspond to the big crunch. 
As in case of CGHS model with dilaton coupled scalars 
in the previous section, we can paste 
the solutions in Figs.7, 8, 9 with the time reversed solution 
at $t=0$. The combined solution would also express the process 
that the universe generated by the big bang disappears by the 
big crunch. Especially the cases of Fig.7 and Fig.8 would 
correspond to $n=1$ case in \cite{XVIII} but the solution 
corresponding to Fig.9 shows rather different behavior.

We can also regard the solution as the 
Kantowski-Sacks Universe which has a $S^1\times S^2$ spatial geometry in (\ref{4}) and the metric (\ref{4dmetric}).
Then the solution of Fig.7 expresses the 
universe where $S^1$ crunches to become a point 
but the radius of $S^2$ diverges in the finite time.
In the solution of Fig.7, the 4d scalar curvature 
diverges positively.
On the other hand, the solutions of Fig.8 and 9 express the 
universe where both of $S^1$ and $S^2$ crunch.
In the solution of Fig.8, the 4d scalar curvature 
diverges negatively at the singularity.
Remarkably, the 4d scalar curvature vanishes and is not 
singular at the singularity in the solution of Fig.9. That 
indicates to a possibility of constructing new versions 
of topologically non-trivial inflationary Universe.

\newpage

\noindent
{\bf Figure Captions}
\begin{flushleft}
Fig.1 Conformal time $t$ (horizontal line) versus 
$\rho$ (vertical line) for 
$t_0=1$ (solid line), $t_0=0$ (dashed line) 
and $t_0=-1$ (dot-dashed line). \\
Fig.2 Cosmological time $\hat t$ (horizontal line) versus 
$\rho$ (vertical line) for 
$t_0=1$ (solid line), $t_0=0$ (dashed line) 
and $t_0=-1$ (dot-dashed line). \\
Fig.3 $t$ versus $\phi$ (solid line) and $\rho$ (dashed line)
for $N=1$, $a=1$, $\lambda^2=-1$ and $t_0=-0.823699$. \\
Fig.4 $t$ versus $\phi$ (solid line) and $\rho$ (dashed line)
for $N=1$, $a=1$, $\lambda^2=1$ and $t_0=-0.141791$. \\
Fig.5 $t$ versus $\phi$ (solid line) and $\rho$ (dashed line)
for $N=1$, $a=1$, $\lambda^2=0$ and $t_0=-0.523121$. \\
Fig.6 $t$ versus $\phi$ (solid line) and $\rho$ (dashed line)
for $N=100$, $a=1$, $\lambda^2=1$ and $t_0=-0.487264$. \\
Fig.7 $t$ versus $\phi$ (solid line), $\rho$ (dashed line) 
and $R_{4d}$ for $N=1$, $a=0$, $\lambda^2=1$. \\
Fig.8 $t$ versus $\phi$ (solid line), $\rho$ (dashed line) 
and $R_{4d}$ for $N=1$, $a=0$, $\lambda^2=1$. \\
Fig.9 $t$ versus $\phi$ (solid line), $\rho$ (dashed line) 
and $R_{4d}$ for $N=100$, $a=1$ and $\lambda^2=1$. \\
\end{flushleft}

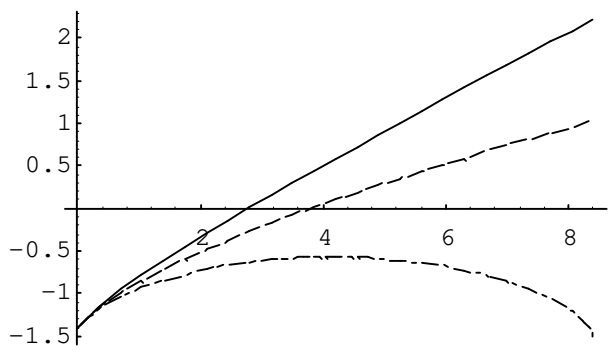
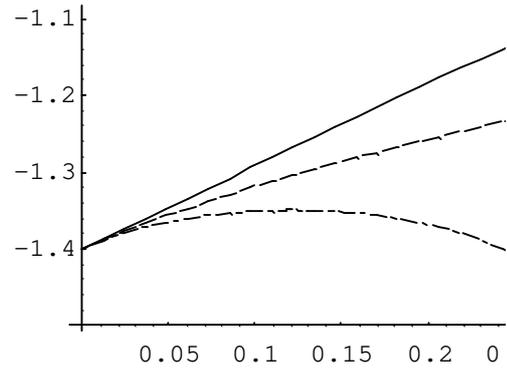
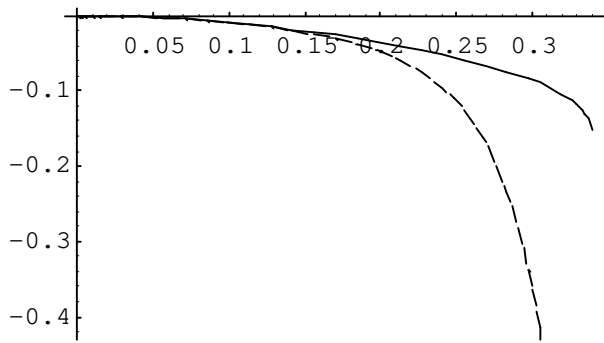
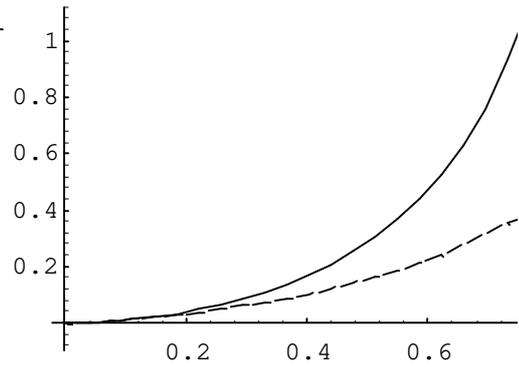
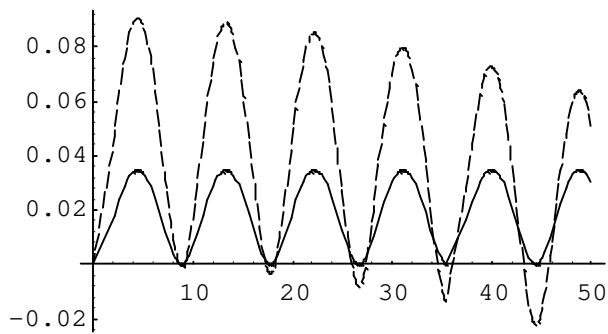
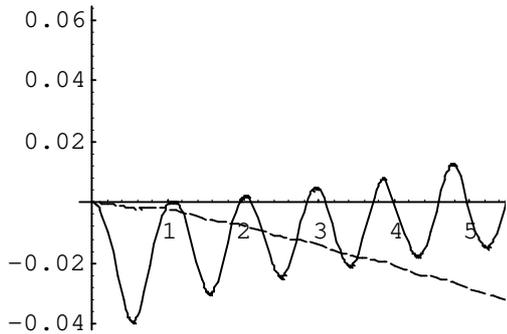
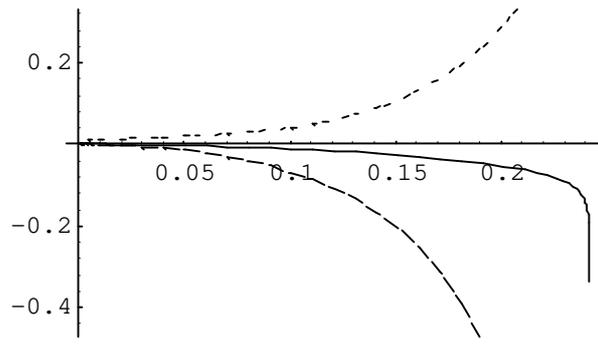
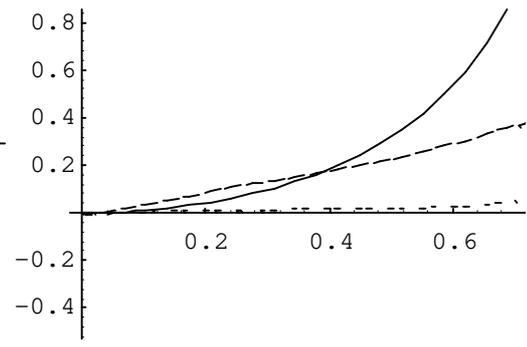
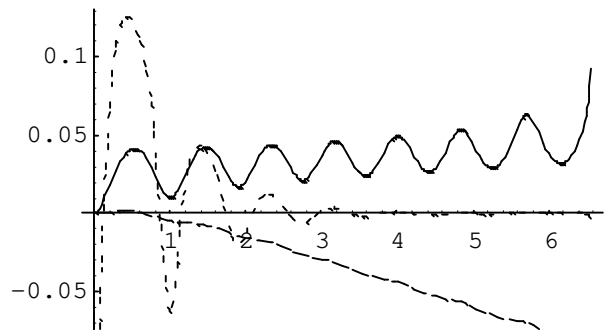
\end{document}